\documentclass[aps,groupedaddress,manuscript]{revtex4}
\usepackage{xcolor}
\usepackage{graphicx}
\usepackage{latexsym}
\usepackage{bm}
\usepackage{amssymb}

\begin{document}

%\draft command makes pacs numbers print
%\draft\footnote{}

%\preprint{to be submitted to PRL}

\title{Agreement at last: an experimental and theoretical study on the single ionization of helium by fast proton impact}
%\title{High-resolution 3D imaging of electron emission in single ionization of helium by fast proton impact}
%\title {Single ionization of helium by fast proton impact - on the search for the influence of coherence properties of the projectile}

\author{H.~Gassert$^1$}
\author{O.~Chuluunbaatar$^{2,3}$}
\author{M.~Waitz$^{1}$}
\author{F.~Trinter$^{1}$}
\author{H.-K.~Kim$^{1}$}
\author{T.~Bauer$^{1}$}
\author{A.~Laucke$^{1}$}
\author{Ch.~M\"{u}ller$^{1}$}
\author{J.~Voigtsberger$^{1}$}
\author{M.~Weller$^{1}$}
\author{J.~Rist$^{1}$}
\author{M.~Pitzer$^{1}$}
\author{S.~Zeller$^{1}$}
\author{T.~Jahnke$^{1}$}
\author{L. Ph. H.~Schmidt$^{1}$}
\author{J.~B.~Williams$^{4}$}
\author{S. A. Zaytsev$^5$}
\author{A. A. Bulychev$^2$}
\author{K. A. Kouzakov$^{6}$}
\author{H.~Schmidt-B\"{o}cking$^{1}$}
\author{R.~D\"{o}rner$^{1}$}
\author{Yu.~V.~Popov$^{2,7}$}
\author{M.~S.~Sch\"{o}ffler$^{1}$}
\email{schoeffler@atom.uni-frankfurt.de}

\affiliation{$^1$ Institut f\"ur Kernphysik, Universit\"at Frankfurt, Max-von-Laue-Str. 1, 60438 Frankfurt, Germany}
\affiliation{$^2$ Joint Institute for Nuclear Research, Dubna, Moscow region 141980, Russia}
\affiliation{$^3$ Institute of Mathematics, National University of Mongolia, UlaanBaatar, Mongolia}
\affiliation{$^4$ Department of Physics,University of Nevada, Reno, Nevada 89557, USA}
\affiliation{$^5$ Department of Physics, Pacific State University, Tikhookeanskaya 136, Khabarovsk 680035, Russia}
\affiliation{$^6$ Faculty of Physics, Lomonosov Moscow State University, Moscow 119991, Russia}
\affiliation{$^7$ Skobeltsyn Institute of Nuclear Physics, Lomonosov Moscow State University, Moscow 119991, Russia}

\vskip 5mm

\date{\today}

\vskip 5mm

\begin{abstract}
Even though ion/atom-collision is a mature field of atomic physics great discrepancies between experiment and theoretical calculations are still common. Here we present experimental results with highest momentum resolution on single ionization of helium induced by 1\,MeV protons and compare these to different theoretical calculations. The overall agreement is strikingly good and already the first Born approximation yields good agreement between theory and experiment. This has been expected since several decades, but so far has not been accomplished. The influence of projectile coherence effects on the measured data is shortly discussed in line with an ongoing dispute on the existence of nodal structures in the electron angular emission distributions.
\end{abstract}


\maketitle

% body of paper here

The ionization dynamics of atoms and molecules are subject to investigation using all kinds of ionizing projectiles. While comparably young fields as ionization induced by synchrotron light and lasers (and even electron impact ionization) show better and better agreement between experiments and theory, for ion-atom collisions - the most mature of these fields - theory and experiment show the biggest unexplained discrepancies, still.

Such disagreement between the most advanced theories and experiments are particularly surprising in a regime where perturbation theory should work best, i.e. for collisions in which projectiles are fast compared to the electron orbital velocity and projectile charges are low: with increasing projectile velocity and decreasing perturbations, fewer terms of the Born series have to be taken into account for an appropriate description. For very fast projectiles, when the interaction times can be as short as sub-attoseconds, it has been expected for a long time that already the first Born approximation should match the experiments perfectly. The regions in phase space most sensitive to deficiencies of a theory are those where the amplitudes of the dominant process vanishes. This leads to nodes in the predicted electron angular distributions. In these cases many delicate mechanisms of interaction can manifest themselves. These can be contribution from higher order terms in the Born series, with prominent examples being the turn-up effect in (e,2e) electron momentum spectroscopy of atoms \cite{Brion98}, effects of the photon momentum in ($\gamma$,2e) \cite{Amusia75, Weber13} and the dipole Cooper minimum in photoionization \cite{Ditchburn28, Cooper}, etc. 

The experimental results of Schulz {\it et al.} published about a decade ago \cite{Schulz2003Nature} are another example of this kind and have launched an avalanche of discussions since their publication. We briefly recall the essence of that experiment: a helium atom was singly ionized by a 100\,MeV/u C$^{6+}$ projectile. The momentum transfer $q$ was fixed to 0.75\,a.u. and the electron energy to 6.5\,eV. For these collision conditions a double lobe structure of a so-called \emph{binary} and \emph{recoil peak} with a distinct node in between had been expected. The three dimensional emission pattern of the electron, however, showed a node only in the plane spanned by the momentum transfer and the projectile momentum. In this collision plane experiment and theory agreed quite well. In the plane perpendicular to the momentum transfer the expected node was filled and theory and experiment showed severe disagreement. While even the most advanced theories to date still predict a node \cite{Madison, Harris, Walters1, Kouzakov12}, further experiments showed a similar behavior of a (partly) filled node \cite{Fischer2004jpb,Schulz2005nimb,Wang2012jpb}. The authors of \cite{Olson2005prl, Kouzakov12} suggested, that the origin of the discrepancies observed in Schulz' experiment is a result of insufficient momentum resolution, which however was refuted by Schulz et al. \cite{comment}. Since then the question remains, whether this disagreement of theory and experiment is due to fundamental reasons thus indicating a general problem of the field of ion-atom collision. The present work tries to shed light onto this subject by presenting the results of an analogous experiment peformed with 1~MeV protons at a similar perturbation strength of $Z_p/v_p$=0.16\,a.u. (as compared to 0.1\,a.u. in \cite{Schulz2003Nature}). With our experimental setup we achieved the highest resolution ever reported in such an ionizing ion-atom collision to definitely rule out possible experimental sources for a disagreement between theory and experiment and to provide a benchmark data set for future calculations in this field. In the following we present the experimental apparatus and the performed calibrations in more detail, as this seems to be necessary given the ongoing dispute on the influence of the experimental resolution on previous results. Atomic units are used throughout.

\section*{Experiment}

The experiment was performed at the Institut f\"ur Kernphysik at the University of Frankfurt using a Van de Graaff accelerator and the well established cold target recoil ion momentum spectroscopy technique (COLTRIMS) to measure the momentum vectors of all charged fragments created in the reaction \cite{Doerner2000pr} in coincidence. A 1\,MeV proton beam from the accelerator (defining the $z$-direction of the laboratory coordinate frame) was collimated using a set of variable slits with an opening of 0.5 $\times$ 1\,mm$^2$ ($x \times y$). At 3.8\,m downstream a second set of slits with an opening of 0.5 $\times$ 1.5\,mm$^2$ was placed. An oscillating electric field ($\approx$150 V/cm), applied on a 30 cm long set of deflector plates 1 m behind the first collimation was used to chop the beam (for more details see \cite{Weber2000jpb}) into buckets of 1 ns length at a repetition rate of 2\,MHz. The projectile beam was crossed at right angle with a supersonic He gas jet (defining the $y$-direction of the coordinate frame). The jet was created by expanding precooled (40\,K) He gas with a stagnation pressure of 2 bar through a 30\,$\mu m$ nozzle, resulting in a speed ratio larger than 100, a target density of $2\times10^{11}$ atoms/cm$^2$ and a jet diameter of 1.5\,mm at the intersection region. Accordingly, a momentum resolution in expansion direction of $\Delta K_{p,y}$=0.1\,a.u. could be achieved. Ions and electrons created in the intersection volume of the projectile and target beam are accelerated by a weak electric field (in $x$-direction) of $E=6.8$\,V/cm towards two position- and time-sensitive detectors. The electron arm of the spectrometer was employed in a time-focusing geometry \cite{Wiley1955rsi} in order to increase the momentum resolution. To reduce the diminishing influence of the extended intersection volume on the experimental resolution even further, the ion side of the spectrometer was designed as a time- and space-focusing geometry (see \cite{doerner1997nimb, schoeffler2009pra, kim2012pra}). More details on this set-up can be found in \cite{Schoeffler2005jpb, schoeffler2009pra2}. The charged fragments were detected using multichannel plate (MCP) detectors with delay line anodes for position read out \cite{Jagutzki2002nima}. Hexagonal anodes \cite{Jagutzki2002ieee} were used with diameters of 120\,mm (electrons) and 40\,mm (ions), respectively. The hexagonal approach allows for an automatic correction of nonlinearity effects, resulting in a dramatic improvement of the overall linearity and local position resolution to values of 100\,$\mu m$ (FWHM). A weak magnetic field of 7.5 Gauss was superimposed parallelly to the electric field to guide the electrons towards the detector \cite{Moshammer1996nim}. From the impact position on the detectors, the time of flight [TOF(He$^+$)=18\,$\mu$s], the spectrometer geometry and the values of the $\vec{E}$-/$\vec{B}$-fields, the momentum vectors of electron and ion have been derived. While the projectile momentum vector has not been measured directly, the excellent momentum resolution of the electron and ion allows deducing it based on momentum conservation. As the accuracy of the angular distributions presented later is extremely sensitive on the exact calibration of the setup, we discuss this procedure in the following in more detail. 

For the ion momentum calibration a scheme similar to that in \cite{Schoeffler14} was applied. Accordingly, we investigated the electron transfer of He$^+$ projectiles at 400 keV impact energy, colliding with a He target in a calibration measurement. In Fig. 1(a) the He$^+$ momentum distribution is shown in the longitudinal direction. The various peaks in the longitudinal momentum $K_{z,ion}$ correspond to a momentum transfer, which is proportional to the total energy difference $Q$ of the initial and the final states of both, the target and the projectile particle \cite{ali1992prl,Mergel1995prl}. As the electronic states are discrete, the width of the observed line is a direct measure of the momentum resolution in the longitudinal direction $z$. We achieved a value of $\Delta K_{z,ion}$=0.09\,a.u. (FWHM). The transversal components of the measured ion momenta for 1 Mev p+He ionization, showing the rotational symmetry around the beam axis, are presented in Fig. 1(b). Both transversal momentum components ($x$ and $y$), have the same center ($K_{x,ion}=K_{y,ion}$=0) and peak width, confirming the correct calibration of the "time-of-flight" and the "position" direction. The same holds for the electron momenta in $x$- and $y$-direction (not shown here). The longitudinal momenta of electron and He$^+$ are directly related, as long as the electron transversal momentum is negligible ($K_{z,ion}=-Q/v_p+v_{z,e}^2/2v_p-v_{z,e}$). The black line in Fig. 1(c) corresponds to $Q$=-24.6\,eV (ionization potential of helium) and the red line to $Q$=-65\,eV (i.e. an additional excitation of the He$^+$ into n=2) providing an independent cross-check of our calibrations and the overall resolution of our system. Finally the overall energy balance $Q$ can be calculated (see Eq.~(3) in \cite{weber2001prl}) and is shown in Fig. 1(d). The measured value fits well to the ionization energy of helium. The long tail to the left corresponds to those events, where the residual He$^+$ ion is additionally excited.

\begin{figure}[htb]
  \begin{center}
	\includegraphics[width=15cm]{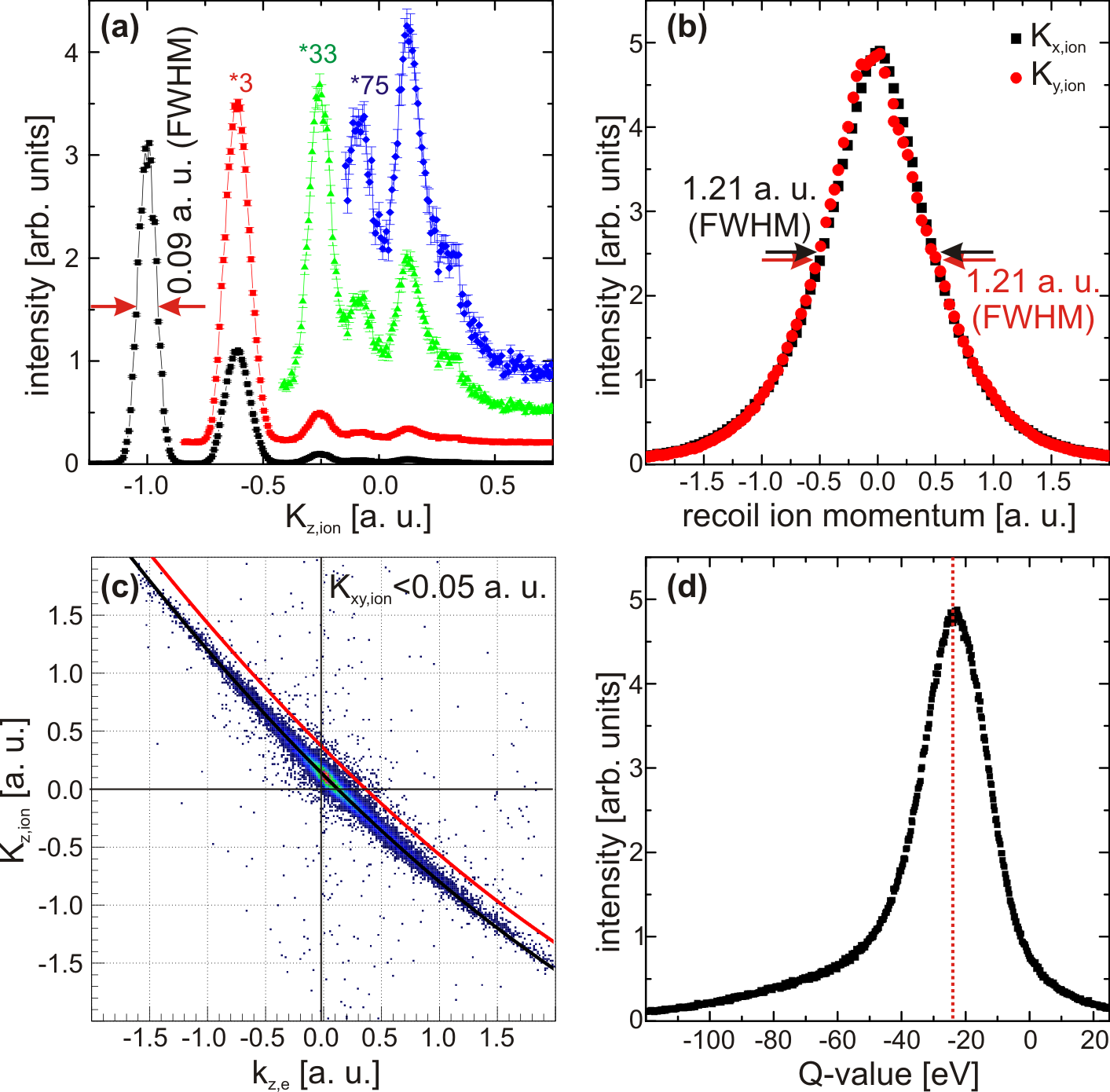}
    \caption{(Color online) (a) He$^+$ longitudinal momentum distribution of the electron transfer reaction: 400 keV ${\rm He}^+_P + {\rm He}_T \rightarrow {\rm He}^0_P(nl) + {\rm He}^+_T(nl)$, with $nl$ being the different final electronic states used for calibration (for more details see \cite{schoeffler2009pra, kim2012pra}). (b-d) 1 MeV p+He ionization data. (b) Transversal He$^+$ momentum distributions in $x$- (black squares) and $y$-direction (red circles); both peaks have the same center of gravity and width. (c) longitudinal momentum ($k_z$/$K_{z,ion}$) of the electron vs. that of the ion for small ion transversal momenta ($K_{xy,ion}<$0.05\,a. u.). The red and black lines are calculations for helium ionization and ionization plus excitation (see main text). (d) Overall energy balance, peaking close to the ionization energy of helium ($Q$=-24\,eV).}
  \end{center}
  \label{fig1}
\end{figure}

\section*{Theory}

In this work, we mainly performed calculations within the plane wave first Born approximation (PWFBA), i.e. when the fast proton is treated as a plane wave both in the initial and final state. The value of the momentum transfer $\vec q=\vec p_i-\vec p_s$ and the energy of the ionized electron $E_e$ are rather small, namely $q=0.75$\,a.u. and $E_e=6.5$\,eV. The law of momentum conservation

$$ 
\vec q = \vec k_e+\vec K_{ion}, \eqno (1) 
$$

illustrates that the velocity of the residual ion $K_{ion}/(m_N+1)$ is negligible, considering its comparably high mass ($m_N\approx 4m_p=7344.6$\,a.u.). This allows us to assume it to be at rest during the reaction and to choose it as a center of the laboratory coordinate system. 

The matrix element is given by:

$$
T_{fi}=\sqrt{2}Z_p\int d^3R d^3r_1 d^3 r_2 \Psi^{-*}_f(\vec R, \vec r_1,\vec r_2; \vec p_s,\vec k_e)\Phi_i(\vec r_1,\vec r_2)\
$$

$$
\times e^{\imath\vec R\vec p_i}\left[\frac{2}{R}-\frac{1}{|\vec R-\vec r_1|}-\frac{1}{|\vec R-\vec r_2|}\right]. \eqno (2)
$$

The factor $\sqrt{2}$ accounts for the identity of the electrons labeled as 1 and 2. The function $\Phi_i$ describes the He atom in its initial (ground) state, and $\Psi^{-}_f$ is the wave function of the full Hamiltonian with the final boundary conditions describing the singly ionized state. $\vec R$ corresponds to the distance between the heavy particles within the model of immovable nucleus.

The energy conservation law

$$ 
E=\frac{p^2_i}{2m_p}+\varepsilon_0^{He}=\frac{(\vec p_i-\vec q)^2}{2m_p}+\varepsilon_0^{He^+}+\frac{k^2_e}{2}+\frac{K^2_{ion}}{2(m_N+1)} \eqno (3) 
$$

allows to obtain the longitudinal and transversal components of the momentum transfer with respect to the incident proton momentum, $q_z=(-\varepsilon_0^{He}+\varepsilon_0^{He^+}+E_e)/v_p=0.18$\,a.u.
and $q_\perp\approx m_p v_p\theta_s$=0.73\,a.u., respectively, where $\theta_s$ is the scattering angle of the proton. In Eq.~(3), we neglect the $q^2/2m_p$ and $K^2_{ion}/2(m_N+1)$ terms, in line with our frozen nucleus approximation.

%The triple differential cross section is calculated on the basis of the formula

%$$ 
% \frac{d^3\sigma}{dE_ed\Omega_ed\Omega_s}= k_e\frac{m^2_p}{(2\pi)^5}|T_{fi}|^2. \eqno (4) 
% $$

The final state of the reaction contains three charged particles in the continuum, namely $p$, $e$ and the He$^+$ ion. In general, the Dollard asymptotic conditions \cite{Dollard} must be taken into account in such a case. However, since the proton energy is high enough, these conditions bring only a very minor effect and, hence, can safely be neglected.

First of all, we choose simple models for the initial and final state. The final state wave function is described by: $\Psi^-_f(\vec R, \vec r_1,\vec r_2; \vec p_s,\vec k_e)=\exp(\imath\vec R\cdot\vec p_s)\Phi^-_f(\vec r_1,\vec r_2; \vec k_e)$. In turn, the final He state with one electron in the continuum $\Phi^-_f(\vec r_1,\vec r_2; \vec k_e)$ is treated as a product of a hydrogen-like He$^+$ ground state wave function and the wave function of the ejected electron in the Coulomb field of the residual He$^+$ ion. The helium ground state $\Phi_i$ is presented by two trial functions: a weakly correlated Roothaan-Hartree-Fock wave function (RHF) \cite{RHF} (we call this FBA for brevity and here both the ground and final helium states are loosely correlated) and a strongly correlated wave function of Ref.~\cite{Chuka} (we call this model c-FBA, and here the helium ground state is highly correlated, but its single continuum final state is still loosely correlated).

We also performed numerical calculations of $\Phi_i$ and $\Phi^-_f$ within the J-matrix approach \cite{Knyr,Knyr2} (j-FBA, where both the ground and final state of helium are highly correlated).

Another model, which we used for estimations is the eikonal wave Born approximation (EWBA). EWBA is a variant of the well-known continuum distorted wave (CDW) approach, and the way of obtaining the phase-factor below was pointed out in \cite{kim2012pra}. Within this approximation we obtain $\Psi^-_f(\vec R, \vec r_1,\vec r_2; \vec p_s,\vec k_e)=\exp(\imath[\vec R\cdot\vec p_s-\eta f(\vec R,\vec r_1,\vec r_2)])\Phi^-_f(\vec r_1,\vec r_2; \vec k_e)$, with

$$
f(\vec R,\vec r_1,\vec r_2)=
$$

$$
\ln\left[\frac{[v_p|\vec R-\vec r_1|+\vec v_p\cdot(\vec R-\vec
r_1)]\ [v_p|\vec R-\vec r_2|+\vec v_p\cdot(\vec R-\vec
r_2)]}{[v_pR+\vec v_p\cdot\vec R]^2}\right]. \eqno (4)
$$

Here we used $v_p=p_i/m_p$=6.33\,a.u., $\eta=Z_p/v_p$, $Z_p$=1, and the proton mass $m_p$=1836.15\,a.u. For calculating the integral (2), we employed the method of 9D integration described in \cite{Schoeffler14}. Also for estimations, calculations in the second Born approximation (PWSBA) were performed in addition to PWFBA, where we used the closure approximation for the Green's function. The details can be found in \cite{Kouzakov12}.

\section*{Results and Discussion}

\begin{figure}[htbp]
  \begin{center}
    \includegraphics[width=15cm]{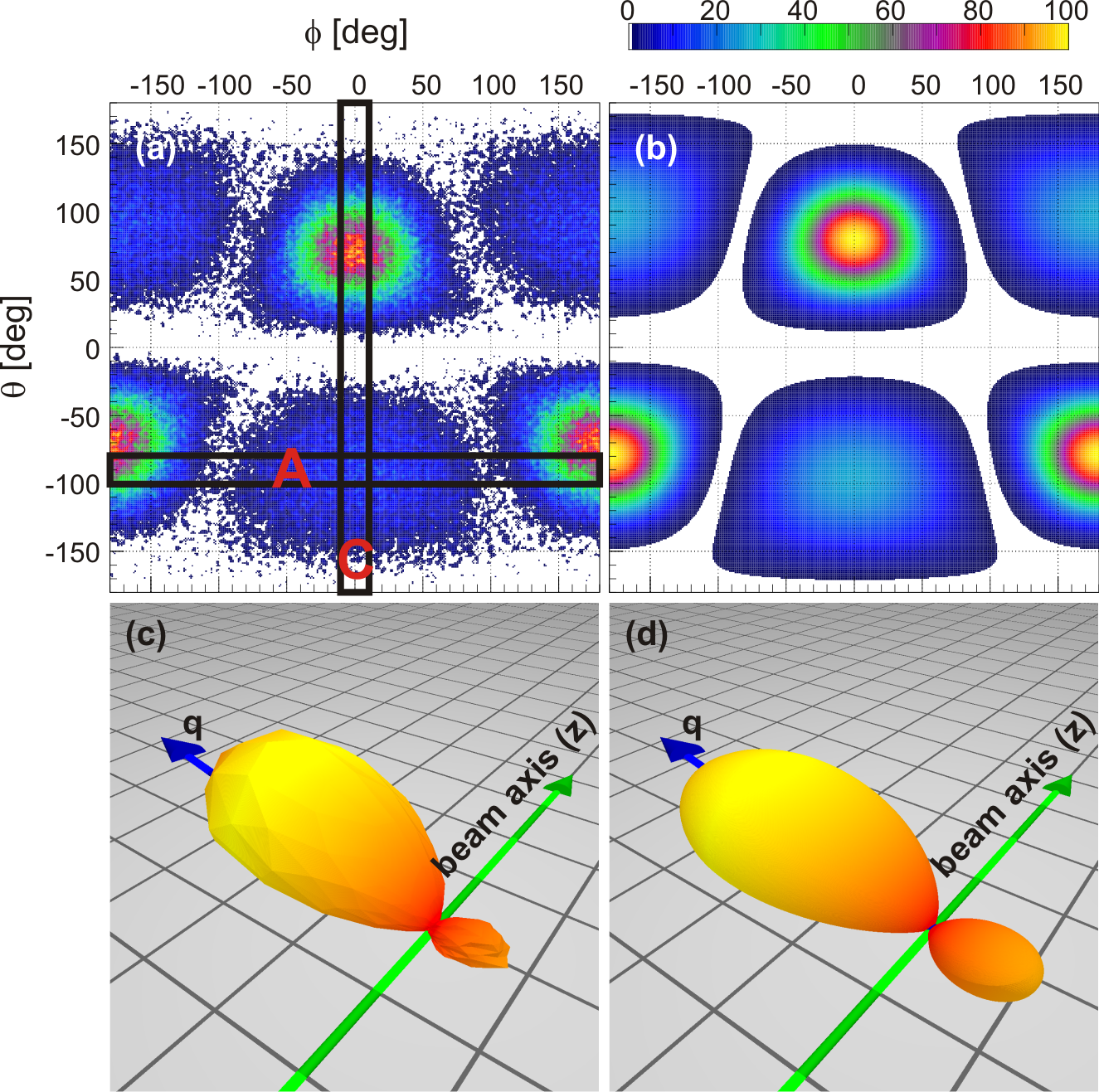}
    \caption{(Color online) Electron angular distributions for a fixed energy of $E_e$=6.5$\pm3.5$\,eV and momentum transfer of $q$=0.75$\pm$0.25\,a.u. (a) Experimental result and (b) theoretical distribution based on the FBA calculations. The areas marked as "A" and "C" correspond to the so-called \emph{azimuthal plane} and the \emph{coplanar geometry}. (c) and (d) depict 3D-representations of the contour plots (a) and (b). The blue arrow indicates the direction of $q$ and the green arrow the initial beam axis (z). The experimental data shown in (c) are mirrored at $\phi$=0 to reduce statistical fluctuations.}
  \end{center}
  \label{fig2}
\end{figure}

The data are presented in a coordinate frame, defined by the initial projectile propagation direction $z$ and the momentum transfer to the projectile $q$. The azimuthal angle $\phi$ is defined around $z$, while the corresponding polar angle is $\theta$. The theoretical results have been convoluted in 2D ($\phi$ and $\theta$) with an angular resolution of 5 degree. In Fig. 2 the experimental electron angular distribution (a) and a theoretical calculation based on PWFBA (b) are shown for a fixed electron energy of $6.5\pm3.5$\,eV and a total momentum transfer of $q$=0.75$\pm$0.25\,a.u. In (c) and (d) these angular distributions are shown in a 3D-representation with the blue arrow being the direction of momentum transfer $q$ and the green arrow being the beam axis $z$. A strongly pronounced node between the forward emitted large binary and the smaller backward emitted recoil peak are clearly visible in theory and experiment. We emphasize that this node is not filled in any direction.

The angular distribution integrated over $\phi$ as function of $\theta$ is shown in Fig. 3(a). For a quantitative comparison the experimental data are normalized to the integral of the c-FBA calculations. The FBA calculation has a slightly lower (7~\%) and the j-FBA theory a slightly larger (8~\%) total cross section. Fig. 3(c) shows the electron angular distribution in the coplanar geometry, [region marked as "C" in Fig. 2(a) and (b)], also known as the \emph{scattering plane}. To illustrate the good agreement between experiment and theory, the data are presented on a logarithmic scale [and as polar plot in Fig. 3(d)]. At $\theta$=0$^\circ$ the node, separating the forward pointing binary and the backward recoil peak, is clearly visible and is perfectly matched by the experimental data. As can be seen also in the polar plot, the experimental binary lobe matches best the correlated PWFBA (c-FBA) wave function (see Fig. 3(d)). While all the first Born calculations peak at $\theta_e$=79$^\circ$, which is the direction of the momentum transfer, the experimental peak is located at $\theta_e$=73$^\circ$. We attribute this deviation to missing higher order interaction terms, which shift the experimental binary and recoil peak forward. Indeed, our numerical estimations within EWBA and PWSBA for the angular domain of the binary peak show an angular shift of approximately 3$^\circ$ towards the experiment. We furthermore find that the minimum between recoil and binary peak is similarly pronounced for all combinations of $q$ and $E_e$ (not shown); only the ratios between binary and recoil peak vary, as well known from (e,2e) experiments \cite{Ehrhardt1969prl,Duerr2006jpb}.

\begin{figure}[htb]
  \begin{center}
    \includegraphics[width=15cm]{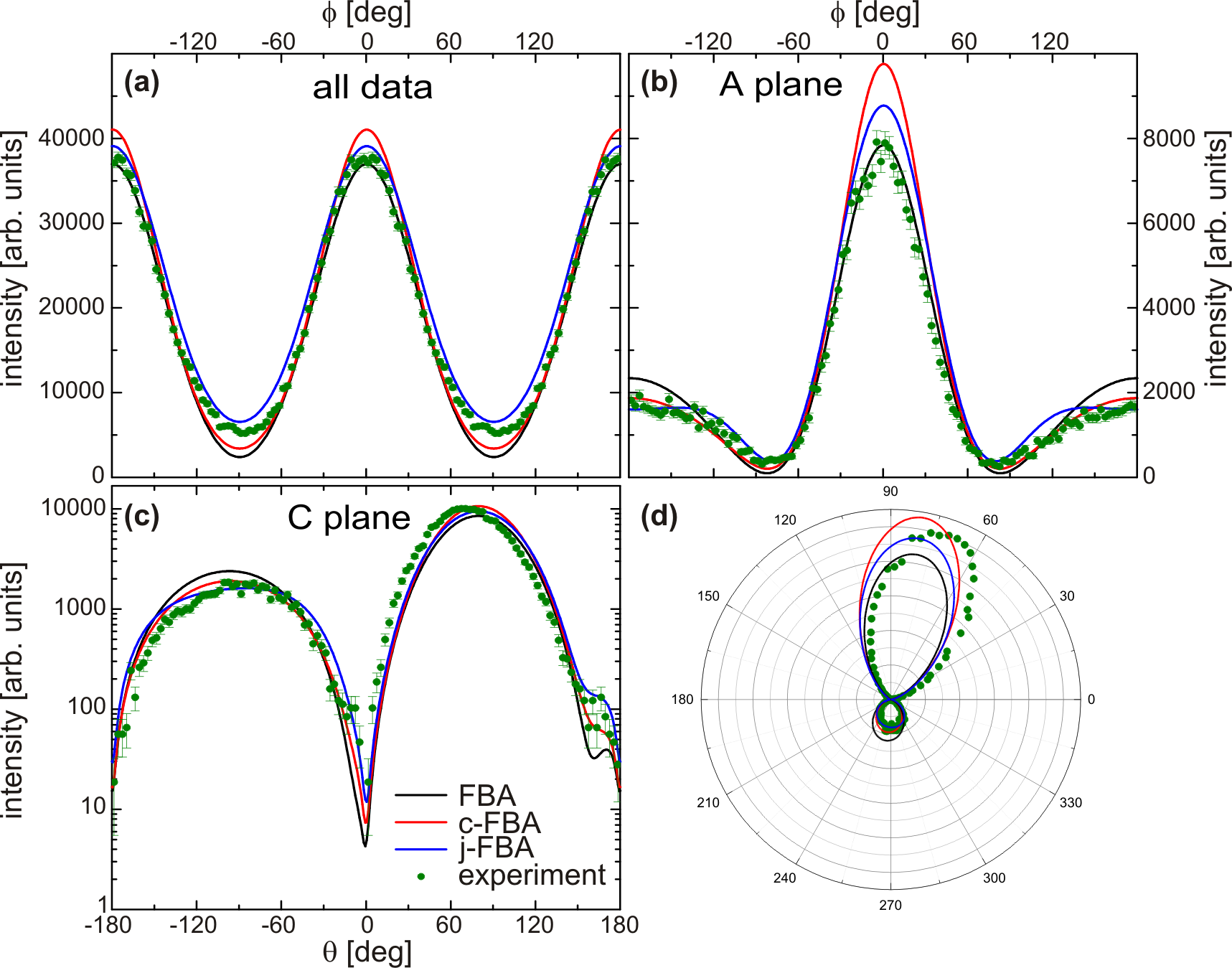}
    \caption{(Color online) Experimental (green dots) and calculated (FBA=black line, j-FBA=blue line, c-FBA=red line) electron angular distribution for $E_e=6.5$\,eV, $q=0.75$\,a.u. in the plane as indicated in Figure 2(a): (a) all data, (b) azimuthal plane, ($\theta_e=90^\circ \pm 10^\circ$), (c+d) coplanar geometry,  ($\phi_e=90^\circ \pm 10^\circ$).}
  \end{center}
  \label{fig3}
\end{figure}

Figure 3(b) shows the angular distribution in the azimuthal plane, marked by region A in Fig. 2(a). The simple FBA agrees best with the experimental data. It has to be noted that due to the tilt of the binary lobes between theory and experiment, the direct comparison does not reflect its quality in this plane. The deviation is simply a result of the different tilts. %The same argument holds for the so-called perpendicular plane, which Schulz and co-workers have addressed in numerous publications \cite{Schulz2003Nature,Wang2012jpb}. Because of the mismatch in the scattering plane, statistics from the binary peak would be accumulated in the distribution of the perpendicular plane, as can also be seen in Fig. 3(c) and (d). 

\begin{figure}[htbp]
  \begin{center}
    \includegraphics[width=15cm]{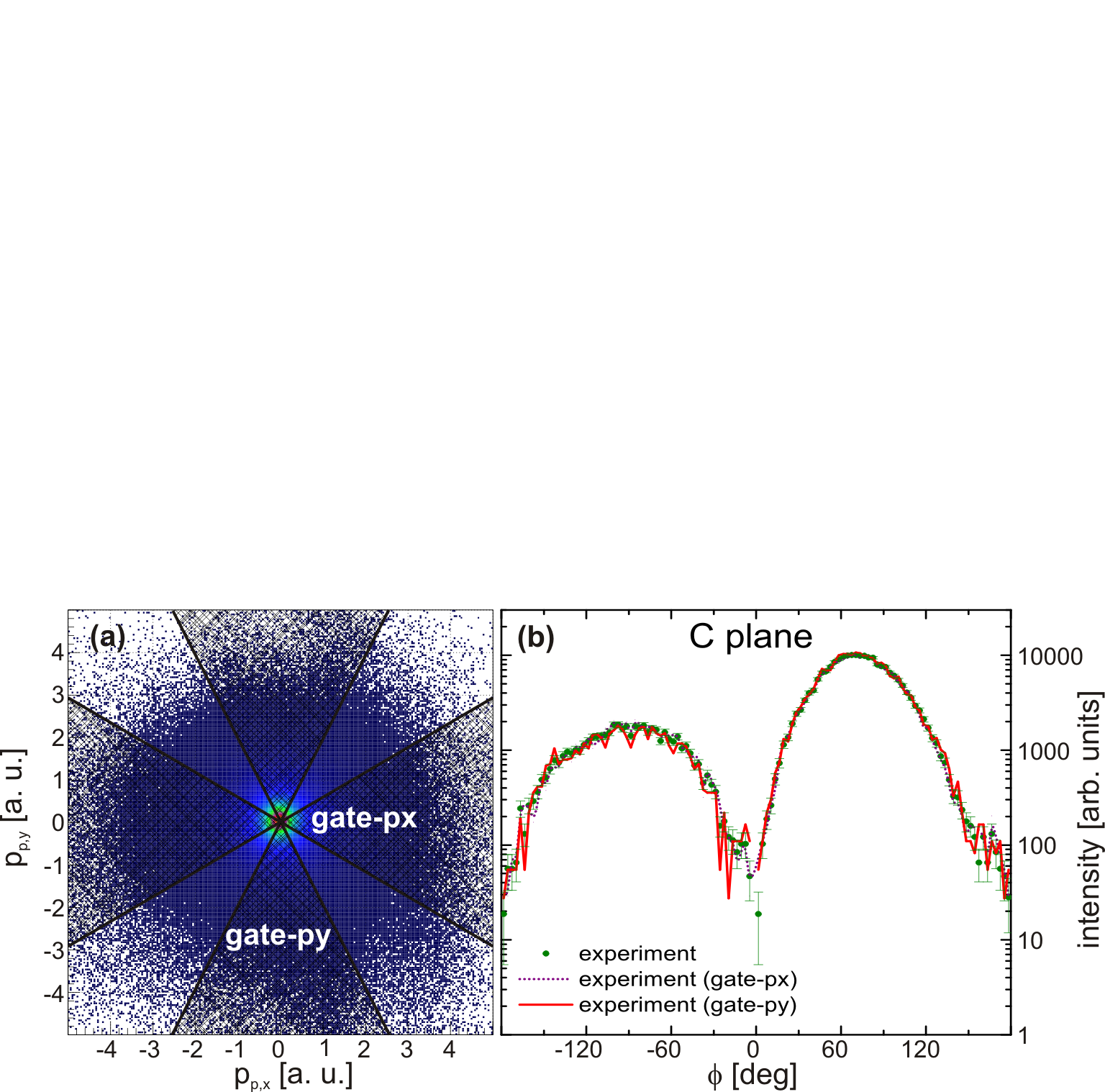}
    \caption{(Color online) (a) projectile momenta in the laboratory $xy$-plane, gated on $\phi^{lab}_p=0\pm30^\circ$ and $\phi^{lab}_p=180\pm30^\circ$ (gate-px) and respectively $\phi^{lab}_p=90\pm30^\circ$ and $\phi^{lab}_p=-90\pm30^\circ$ (gate-py). (b) The experimental electron angular distribution in the C-plane gated and ungated.}
  \end{center}
  \label{fig4}
\end{figure}

In conjunction with the discussion of the strong theory-experiment deviation connected to the C$^{6+}$-experiments of Schulz et al. \cite{Schulz2003Nature}, we discuss the concept of so-called projectile coherence in general and for our present data. Schulz et al. claim a transversal coherence of their projectile beam of $\Delta_x=10^{-3}$\,a.u. \cite{Wang2012jpb}. In the same publication \cite{Wang2012jpb} the authors claim that a coherence length of $\Delta_x$=0.25\,a.u. is already sufficient to yield an incoherent beam, while $\Delta_x$=4\,a.u. corresponds to a coherent beam. For the case of a coherent beam they observe an evidence for a node between the binary and the recoil peak, which then vanishes as the beam becomes incoherent. A part of the filling of the node is attributed to the limited momentum resolution of that experiment. As our projectile beam was rectangularly collimated, a divergence of $\delta_x$=0.26\,mrad in the $x$-direction and $\delta_y$=0.65\,mrad in the $y$-direction was achieved. According to \cite{Wang2012jpb}, this corresponds to a transversal coherence of 2.1\,a.u. in $x$-direction and 0.8\,a.u. in $y$-direction, assuming that $\Delta_{x/y}=\delta_{x/y}\times\lambda$, with $\lambda$ being the de Broglie wavelength of the projectile. As the projectile beam has a different transversal coherence in x- and y-direction in the laboratory frame, we search for its potential influence on the electron angular distribution: in Fig. 4(a) the transverse projectile momenta are shown. Additionally two area-gates, selecting whether the momentum transfer occurred in the $x$-direction or $y$-direction are depicted. Employing these gates, the electron angular distribution is plotted for the coplanar geometry in Fig. 4(b). Within the error bars we do not observe any difference in the emission patterns obtained for gate-px and gate-py. 

In conclusion we have performed a high resolution experiment on electron emission in fast proton helium collisions. Our data are in full agreement with the expectations from standard scattering theory with a deep node between binary and recoil peak in all directions. We do not find any indications for an influence of a possible reduced coherence of the projectile on the ionization process.

\section{Acknowledgments}

M.S. gratefully acknowledges financial supported from the Deutsche Forschungsgemeinschaft (DFG) under Grant No. SCHO 1210/2-1. The research of three of the authors (K. A. K., Yu. V. P., and A. A. B) is supported by the Russian Foundation of Basic Research (Grant No. 14-01-00420-a). O. Ch. acknowledges support within the Hulubei-Meshcheryakov programme JINR-Romania. All calculations were performed at the Moscow State University Research Computing Centre (supercomputers Lomonosov and Chebyshev) and the Joint Institute for Nuclear Research Central Information and Computer Complex. We acknowledge many years of fruitful discussion with Michael Schulz and Daniel Fischer.

\bibliographystyle{unsrt}

\end{document}